\documentclass[oupdraft]{bio}

\usepackage{url}
\usepackage{outlines}
\usepackage{multirow}
\usepackage{tabularx}
\usepackage{appendix}
\usepackage{algorithm}
\usepackage[noend]{algpseudocode}
\usepackage{setspace}
\usepackage{booktabs}
\usepackage{color}
\usepackage{pdflscape}

\newcommand{\Ourmethod}{\textsc{TrialMDP}}
\newcommand{\Githubrepo}{\url{https://github.com/dpmerrell/TrialMDP}}
\newcommand{\Analysisrepo}{\url{https://github.com/dpmerrell/TrialMDP-analyses}}

\DeclareMathOperator{\argmax}{argmax}
\DeclareMathOperator{\Beta}{Beta}
\DeclareMathOperator{\Binomial}{Binomial}
\DeclareMathOperator{\E}{\mathbb{E}}

\newcommand{\pluseq}{\mathrel{+}=}
\newcommand{\minuseq}{\mathrel{-}=}

\newcommand{\tmin}{T_{\text{min}}}

\DeclareMathOperator{\CMH}{CMH}

\newcommand{\primo}{\emph{(i)}}
\newcommand{\secundo}{\emph{(ii)}}
\newcommand{\tertio}{\emph{(iii)}}

\history{Received Month DD, 2021;
revised Month DD, 2021;
accepted for publication Month DD, 2021}

\begin{document}

\title{A Markov Decision Process\\ for Response-Adaptive Randomization in Clinical Trials}

\author{David Merrell$^\ast$, Thevaa Chandereng$^\dagger$, Yeonhee Park$^\ddagger$\\[4pt]
\textit{$^\ast$Department of Computer Sciences, University of Wisconsin - Madison. 
        $^\dagger$Department of Biostatistics, Columbia University. 
        $^\ddagger$ Department of Biostatistics and Medical Informatics, University of Wisconsin - Madison.
}
\\
{dmerrell@cs.wisc.edu, tc3123@cumc.columbia.edu, ypark56@wisc.edu
}
}

\markboth%
{Merrell, Chandereng, Park}
{MDP Trial Design}

\maketitle

\footnotetext{To whom correspondence should be addressed.}

\begin{abstract}
{

In clinical trials, response-adaptive randomization (RAR) has the appealing ability to assign more subjects to better-performing treatments based on interim results. 
The traditional RAR strategy alters the randomization ratio on a patient-by-patient basis; 
this has been heavily criticized for bias due to time-trends. 
An alternate approach is blocked RAR, which groups patients together in blocks and recomputes the randomization ratio in a block-wise fashion; 
the final analysis is then stratified by block. 
However, the typical blocked RAR design divides patients into equal-sized blocks, which is not generally optimal. 

This paper presents \Ourmethod{}, an algorithm that designs two-armed blocked RAR clinical trials.
Our method differs from past approaches in that it optimizes the \emph{size and number of blocks} as well as their treatment allocations.
That is, the algorithm yields a policy that adaptively chooses the \emph{size and composition} of the next block, based on results seen up to that point in the trial.
\Ourmethod{} is related to past works that compute optimal trial designs via dynamic programming.

The algorithm maximizes a utility function balancing \primo{} statistical power, \secundo{} patient outcomes, and \tertio{} the number of blocks.
We show that it attains significant improvements in utility over a suite of baseline designs, and gives useful control over the tradeoff between statistical power and patient outcomes.
It is well suited for small trials that assign high cost to failures.

We provide \Ourmethod{} as an R package: 
\Githubrepo{}.
}{Adaptive randomization; Clinical trial; Dynamic programming; Markov decision process; Reinforcement learning}
\end{abstract}

\section{Introduction}
\label{sec:intro}





\noindent

Randomization is a common technique used in clinical trials to eliminate potential bias and confounders in a patient population. 
Most clinical trials utilize fixed randomization, where the probability of assigning subjects to a treatment group is kept fixed throughout the trial. 
Response-adaptive randomization (RAR) designs were developed due to the captivating benefit of increasing the probability of assigning patients to more promising treatments, based on the responses of prior patients.
A big downside for RAR designs is that the time between treatment and outcome must be short, in order to inform future patients' randomization.

Traditional RAR designs recompute the randomization ratio on a patient-by-patient basis \citep{thall2007practical}, usually after a burn-in period of fixed randomization. 
However, traditional RAR designs have been widely criticized \citep{karrison2003group}.
Traditional RAR designs induce bias due to temporal trends in clinical trials. 
Temporal trends are especially likely to occur in long duration trials. 
Patients’ characteristics might be completely different throughout the trial or even at the beginning and end of the trial \citep{proschan2020resist}. 
However, standard RAR analyses assume that the sequence of patients who arrive for entry into the trial represents samples drawn at random from two homogeneous populations, with no drift in the probabilities of success \citep{proschan2020resist, chandereng2020response}. 
This assumption is usually violated. 
For example, in the BATTLE lung cancer elimination trials \citep{liu2015overview}, more smokers enrolled in the latter part of the trial than at the beginning.

Despite this serious flaw, there is not much literature to address the temporal trend issue in RAR designs.   
Villar et al. explored the hypothesis testing procedure adjusting for covariates for correcting type-I error inflation and the effect on power in RAR designs with temporal trend effects added to the model for two-armed and multi-armed trials \citep{villar_timetrend_2018}.
\cite{karrison2003group} introduced a stratified group-sequential method with a simple example of altering the randomization ratio to address this issue. 
\cite{chandereng_robust_2019} further examined the operating characteristics of the blocked RAR approach for two treatment arms proposed by \cite{karrison2003group}. 
They concluded that blocked RAR provides a good trade-off between ethically assigning more subjects to the better-performing treatment group and maintaining high statistical power. 
They also suggested using a small number of blocks since large numbers of blocks have low statistical power. 
However, \citet{chandereng_robust_2019} designed trials with equal-sized blocks, which is not generally optimal.

Other works formulate adaptive trial design as a Multi-Armed Bandit Problem (MABP), employing ideas that are often associated with reinforcement learning---e.g., sequential decision-making and regret minimization.
These entail sophisticated algorithms, such as Gittins index computations \citep{villar_bandit_2015, villar_gittins_2015} and dynamic programming \citep{hardwick_exact_1995,hardwick_induction_1999,hardwick_optimal_2002}.
These works have important limitations.
The Gittins index approaches of \citeauthor{villar_gittins_2015} assume either \primo{} a fully sequential trial with similar weaknesses to traditional RAR or \secundo{} a blocked trial with equal-sized blocks.
The dynamic programming algorithms of \citeauthor{hardwick_optimal_2002}
yield allocation rules that \primo{} are deterministic, \secundo{} are fully sequential, or \tertio{} assume a blocked trial with a fixed number of blocks.
At the time, \citeauthor{hardwick_optimal_2002}'s approaches were also limited by computer speed and memory, which have improved famously over the years.

This paper presents \Ourmethod{}, an algorithm that designs blocked RAR trials.
\Ourmethod{} is most closely related to the MABP-based approaches mentioned above.
However, it models a blocked RAR trial as a Markov Decision Process (MDP), a generalization of the MABP.
It relies on a dynamic programming algorithm, similar to those of \citeauthor{hardwick_optimal_2002}.
However, our method differs in that it optimizes the \emph{size and number of blocks} as well as their treatment allocations.
That is, the algorithm yields a policy that adaptively chooses the \emph{size and composition} of the next block, based on results seen up to that point in the trial.
The current version of \Ourmethod{} is tailored for two-armed trials with binary outcomes. 
Future versions may permit a more general class of trials.

Our paper has the following structure.
In Section \ref{sec:methods}, we describe our problem formulation and algorithmic solution. 
In Section \ref{sec:results}, we compare \Ourmethod{}'s designs with other designs that have been widely used in clinical trials. 
We use our proposed method to redesign a phase II trial in Section \ref{sec:redesign}.
We discuss \Ourmethod{}'s limitations and potential improvements in Section \ref{sec:discussion}. 
Our Supplementary Materials include appendices that justify some of our mathematical and algorithmic choices.

\section{Proposed method}
\label{sec:methods}

\subsection{Problem formulation}
\label{sec:formulation}

\paragraph{Class of trials. }
In this paper we focus on blocked RAR trials with two arms and binary outcomes.
We label the arms $A$ and $B$ (``treatment'' and ``control'', respectively) and outcomes 0 and 1 (``failures'' and ``successes'').
A trial has access to some number of available patients, $N$.
The trial proceeds in $K$ blocks.
We require that all results from the current block are observed before the next block begins.
Importantly, we allow $K$ to adapt as the trial progresses.
This gives the trial useful kinds of flexibility.
In general, a trial may attain better characteristics if it permits differently-sized blocks.

Let $p_A, p_B$ denote the treatments' success probabilities.  
We assume a frequentist test is performed at the end of the trial, with the following null and alternative hypotheses $\mathcal{H}_0, \mathcal{H}_A$:
$$ \mathcal{H}_0 : \ p_A = p_B \qquad \mathcal{H}_A : \ p_A > p_B  $$
We focus specifically on the one-sided Cochran-Mantel-Haenzsel (CMH) test, which is well-suited for stratified observations; 
in our setting, the strata are blocks of patients.
It has been argued that blocked RAR trials with CMH tests are more robust to temporal trend effects than, e.g., traditional RAR trials with chi-square tests \citep{chandereng_robust_2019}.

\begin{figure}
    \centering
    \includegraphics[scale=0.7]{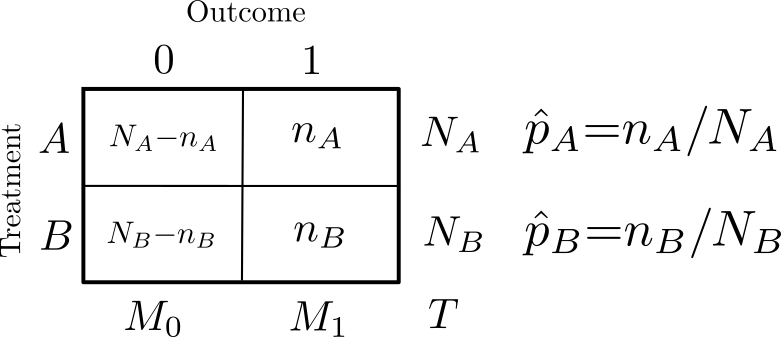}
    \hspace{1em}
    \includegraphics[scale=0.7]{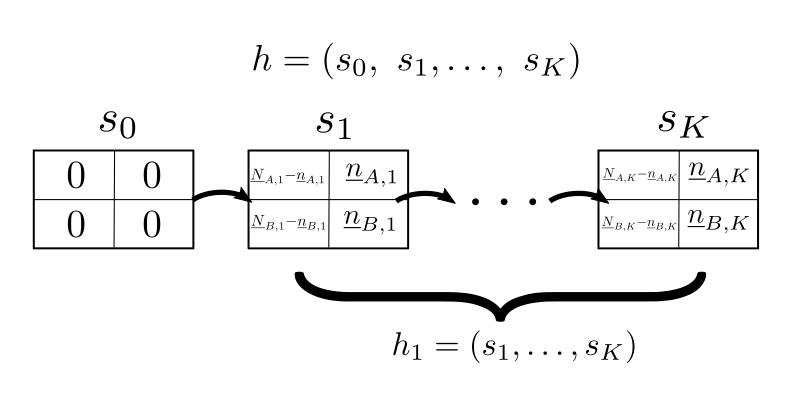}\\
    (A) \hspace{2.75in} (B)
    \caption{(A) Contingency table notation. (B) Trial history notation. A history $h$ is a sequence of cumulative contingency tables, $(s_0, \ldots, s_K)$. A subscript $h_k$ indicates a history's $k^\text{th}$ \emph{suffix}.}
    \label{fig:contingency}
\end{figure}

\paragraph{Notation: tables and histories.}
We establish some notation for clarity.
A $2{\times}2$ contingency table has the following attributes: $N_A$ and $N_B$, the numbers of patients assigned to each treatment; $n_A$ and $n_B$, the numbers of successes for each treatment; $M_0$ and $M_1$, the total numbers of failures and successes; and $T$, the total number of outcomes recorded in the table.
Symbols $\hat{p}_A$, $\hat{p}_B$ represent point estimates of the treatment success probabilities. 
See Figure \ref{fig:contingency} for illustration.

Each block of the trial has its own contingency table with corresponding quantities.
We use a subscript to indicate the block.
For example, the $k^\text{th}$ block of the trial has its own table with quantities $N_{A,k}$, $n_{A,k}$, $T_k$, and so on.

At any point we can summarize the state of the trial in a $2{\times}2$ contingency table, $s$, of \emph{cumulative} results.
That is, $s$ contains \emph{all} of the trial's observations up to that point;
or, put another way, $s$ is the sum of all preceding block-wise contingency tables. 
We typically refer to $s$ as a \emph{state}.
We use an underline to indicate a quantity computed from a state.
For example, after completing $k$ blocks we have quantities $\underline{N}_{A,k} {=} \sum_{i=1}^k N_{A,i}$; \hspace{0.25em}
$\underline{n}_{A,k} {=} \sum_{i=1}^k n_{A,i}$; \hspace{0.25em}
$\underline{\hat{p}}_{A,k} {=} \underline{n}_{A,k}/\underline{N}_{A,k}$; and so on. 

The sequence of states occupied by a trial forms a \emph{trial history} $h {=} (s_0, \ldots, s_K)$, where $s_0$ is always the empty contingency table and $s_K$ always has $\underline{T}_K {=} N$ observations.
We use a subscript to denote the \emph{suffix} of a history.
For example, $h_k {=} (s_{k}, \ldots, s_K)$ is the sequence of states \emph{after} the $k^\text{th}$ block of the trial.
It is useful to think of a history as a random object, subject to uncertainty in the patient outcomes and the values of $p_A$, $p_B$.

\paragraph{Utility function. }
We aim to design blocked RAR trials that balance
\primo{} statistical power and
\secundo{} patient outcomes.
We also recognize that each additional block entails a cost in time and other overhead.
As such, we wish to avoid an excessive number of blocks.
We formalize these goals with the following utility function:
\begin{equation}
    U(h) = V(h) ~-~ \lambda_F {\cdot} F(h) ~-~ \lambda_K {\cdot} K(h)
    \label{eq:utility}
\end{equation}
where $V(h)$ is a proxy for the trial's statistical power; $F(h)$ measures the number of failures; and $K(h)$ is the number of blocks.
This utility function promotes a high statistical power while penalizing failures and blocks.
The coefficients $\lambda_F$ and $\lambda_K$ control the relative importance of patient outcomes and blocks, respectively.

The functions $V$, $F$, and $K$ have the following forms:
$$
V(h) = \frac{1}{N} \sum_{i=1}^K \frac{w_i}{\frac{1}{2}(\underline{\hat{p}}_{A,i} + \underline{\hat{p}}_{B,i}) \frac{1}{2}( \underline{\hat{q}}_{A,i} + \underline{\hat{q}}_{B,i}) }
\hspace{0.35in} 
F(h) = \frac{1}{N}(\underline{N}_{A,K} - \underline{N}_{B,K})(\underline{\hat{p}}_{B,K} -\underline{\hat{p}}_{A,K})
$$
$$ K(h) = K $$
Function $K$ simply returns the number of blocks in the trial history.
Function $F$ quantifies bad patient outcomes (i.e., failures) as a fraction of all patients. It is a function only of the final state, $s_K$, and becomes small when the estimates $\underline{\hat{p}}_{A,K}$ and $\underline{\hat{p}}_{B,K}$ are close.

Function $V$ serves as a proxy for the trial's statistical power.
It is crafted such that maximizing $V$ also maximizes the power of the Cochran-Mantel-Haenzsel test \citep{cochran_cmh_1954} to an acceptable approximation.
Each $w_i = N_{A,i} N_{B,i}/(N_{A,i}{+}N_{B,i})$ is the \emph{harmonic mean} of that block's treatment allocations.
$V$ takes larger values when the allocations are balanced; and when $p_A,p_B$ are close to each other, and far from $\frac{1}{2}$.
The factor $\frac{1}{N}$ makes $V$ consistent across trials with differing sample sizes. 
See Appendix \ref{sec:v_score} of the Supplementary Materials for a more detailed justification of $V$.

\paragraph{Markov Decision Process formulation. }
In our effort to maximize the expected utility (Equation \ref{eq:utility}), we find it useful to model a blocked RAR trial as a \emph{Markov Decision Process} (MDP).
An MDP is a simple model of sequential decision-making.
It consists of an agent navigating a \emph{state space}.
At each time-step, the agent chooses an \emph{action}.
Given the agent's current state and chosen action, the agent \emph{transitions} to a new state and collects a \emph{reward}.
In general the transition is stochastic, governed by a \emph{transition distribution}.
One \emph{solves} an MDP by obtaining a \emph{policy} that maximizes the expected total reward. 
We refer the reader to Chapter 38 of \citeauthor{lattimore_bandit_2020}'s text for detailed information about MDPs \citep{lattimore_bandit_2020}.

\begin{figure}
    \centering
    \includegraphics[scale=0.7]{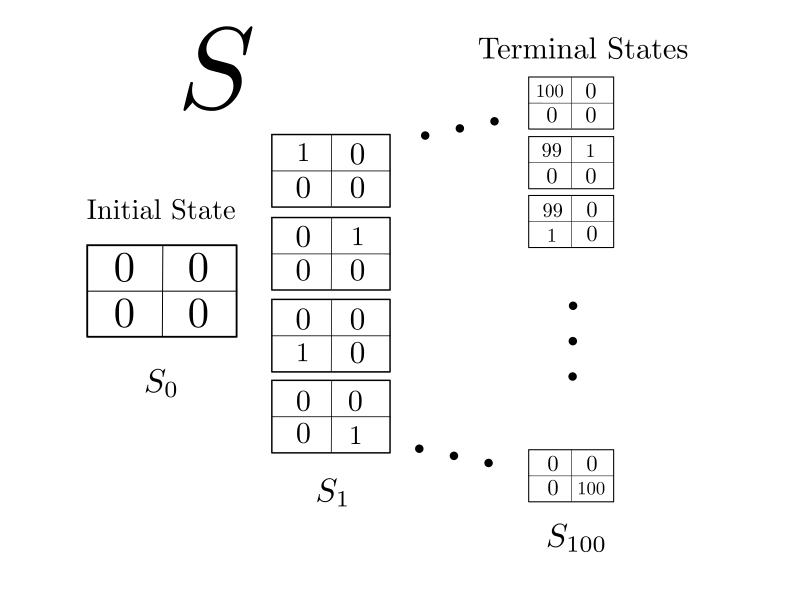}
    \includegraphics[scale=0.7]{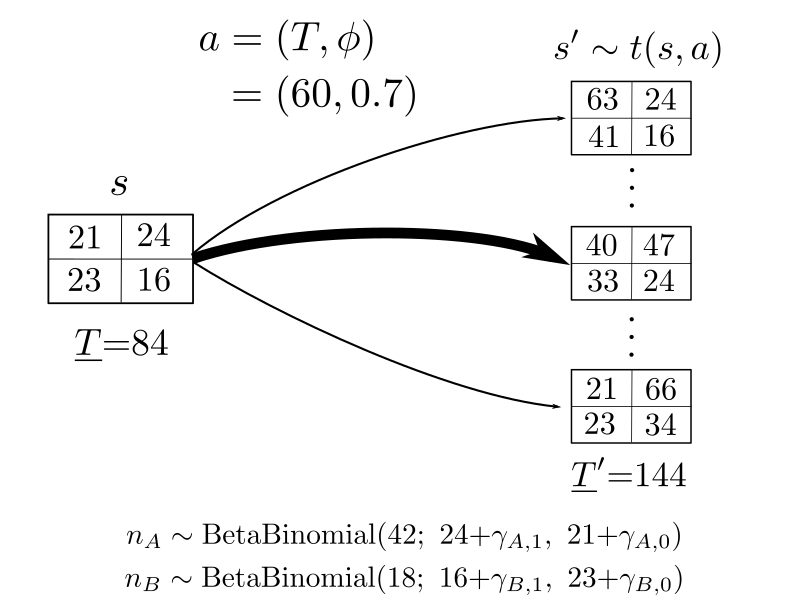}\\
    (A) \hspace{2.5in} (B)
    \caption{(A) State space $S$. At any point, the state of the trial is summarized by a contingency table of all observations. We can order the set of all contingency tables by their numbers of observations, $\underline{T}$. The trial begins with the empty table in $S_0$; the trial ends when it reaches a state in $S_N$ (in this example $N{=}100$). 
    (B) Transition distribution. In this example, current state $s$ and action $a{=}(60,0.7)$ induce a distribution $s^\prime \sim t(s,a)$ for the next state. 
    The next state necessarily has $\underline{T}^\prime {=} 144 {=} 84 + 60$. 
    Its entries are governed by Beta-Binomial distributions, parameterized by the entries of the current contingency table. 
    }
    \label{fig:states}
\end{figure}

We model a blocked RAR trial as an MDP with the following components:
\begin{itemize}
    \item \emph{State space.} 
    In our setting the state space $S$ consists of every possible $2{\times}2$ contingency table with ${\leq}N$ observations. 
    We can order the states by their numbers of observations.
    We let $S_i$ denote the subset of $S$ containing tables with exactly $i$ observations. 
    The trial always begins at the empty contingency table in $S_0$ and terminates at some table in $S_N$.
    The state space grows quickly with $N$, $|S|{=}O(N^4)$. Figure \ref{fig:states}(A) illustrates $S$ for $N{=}100$.
    \item \emph{Actions.} 
    With each block of the trial we choose an action $a{=}(T, \phi)$, the block's \emph{size} and \emph{allocation}. 
    Suppose we have completed $k$ blocks; then $T$ may take any integer value from 1 to $N{-}\underline{T}_k$. 
    The allocation $\phi$ is the fraction of patients assigned to treatment $A$ in this block. 
    We constrain $\phi$ to a finite set of possible values, $\Phi$. 
    For example, $\Phi = \{0.2, 0.3, \ldots, 0.8\} $. 
    Importantly, exactly $T{\cdot}\phi$ patients (rounded to the nearest integer) are assigned to treatment $A$.
    In other words, patients are randomized to treatments ``without replacement.'' 
    Contrast this with other randomized designs---traditional RAR, blocked RAR, etc.---that assign each patient to $A$ with independent probability $\phi$. For example, action $(T{=}60, \phi{=}0.7)$ implies that the next block will treat 60 patients, assigning exactly $T {\cdot} \phi {=} 42$ of them to treatment $A$ and $18$ to treatment $B$. 
    
    We let $\mathcal{A}$ denote the set of all actions,
    and $\mathcal{A}_s$ denote actions available at state $s$.
    \item \emph{Transition distributions.} 
    Given the current contingency table $s_i$ and the chosen block design $a_{i}{=}(T_i,\phi_i)$, the next contingency table $s_{i+1}$ is randomly distributed. 
    This randomness consists of two parts: \primo{} the stochasticity of patient outcomes given the true success probabilities $p_A$ and $p_B$, and \secundo{} our uncertainty about the values of $p_A$ and $p_B$.
    Given the true values for $p_A$ and $p_B$, the numbers of successes $n_{A,i+1}$ and $n_{B,i+1}$ for this block would have Binomial distributions:
    $$ n_{A,i+1}|p_A \sim \Binomial\left(T_i {\cdot} \phi_i,~ p_A\right) \hspace{0.5in} n_{B,i+1}|p_B \sim \Binomial\left(T_i {\cdot} (1-\phi_i), ~p_B\right).$$
    However, we only have imperfect knowledge of $p_A$ and $p_B$, encoded in the entries of the current table $s_i$.
    We use Beta distributions to describe this uncertainty about $p_A$ and $p_B$:
    $$ p_A \sim \Beta(\underline{n}_{A,i} {+} \gamma_{A1},~ \underline{N}_{A,i} {-} \underline{n}_{A,i} {+} \gamma_{A0} ) \hspace{0.4in} p_B \sim \Beta(\underline{n}_{B,i} {+} \gamma_{B1},~ \underline{N}_{B,i} {-} \underline{n}_{B,i} {+} \gamma_{B0} ) $$
    where each $\gamma_{\ast}$ is a smoothing hyperparameter typically set to 1.
    Together, these two sources of randomness assign independent Beta-Binomial probabilities to $n_{A,i+1}$ and $n_{B,i+1}$, which in turn define the distribution for $s_{i+1}$. 
    See Figure \ref{fig:states}(B) for illustration.
    We sometimes use the notation $s_{i+1} \sim t(s_i,a_i)$ to indicate the \emph{transition distribution} for $s_{i+1}$, given $s_i$ and $a_i$.
    \item \emph{Rewards.} 
    Given the current state $s_i$ and the chosen action $a_{i}$, the trial transitions to state $s_{i+1}$ and receives a reward $R(s_i, a_i, s_{i+1})$.
    In an MDP the goal is to maximize expected total reward.
    Recall, however, that our ultimate goal is to maximize the expected utility $U$ (Equation \ref{eq:utility}).
    We craft a reward function $R$ consistent with $U$, as follows:
    \begin{equation}
        R(s_i, a_i, s_{i+1}) = \begin{cases}                             \frac{1}{N}\frac{w_{i+1}}{\frac{1}{2}(\underline{\hat{p}}_{A,i+1} + \underline{\hat{p}}_{B,i+1}) \frac{1}{2}( \underline{\hat{q}}_{A,i+1} + \underline{\hat{q}}_{B,i+1})} ~-~ \lambda_K & s_{i+1} \notin S_N \\[2em]
         \frac{1}{N}\frac{w_{i+1}}{\frac{1}{2}(\underline{\hat{p}}_{A,i+1} + \underline{\hat{p}}_{B,i+1}) \frac{1}{2}( \underline{\hat{q}}_{A,i+1} + \underline{\hat{q}}_{B,i+1})} ~-~ \lambda_K ~-~ \lambda_F {\cdot} F(s_{i+1}) & s_{i+1} \in S_N 
                             \end{cases}
                             \label{eq:reward}
    \end{equation}
    The total reward for a trial history is identical to the utility (Equation \ref{eq:utility}) of that trial history.
    With each block, the reward function produces that block's contribution to the total utility.
    This includes the block's term for $V$; the block's cost $\lambda_K$; and the final failure penalty $F(s_{i+1})$ when $s_{i+1}$ is terminal.
    
    Notice that our particular $R$ is a function only of $a_i$ and $s_{i+1}$.
    We sometimes write $R(a_i, s_{i+1})$ for compactness.
    \item \emph{Policy.} 
    A policy is a function $\pi: S \rightarrow \mathcal{A}$ mapping each state in the MDP to an action.
    In our setting policies are \emph{trial designs}.
    For each state $s_i$ in the trial, a policy dictates the design of the trial's next block: $\pi(s_i) {=}(T_{i+1},\phi_{i+1})$.
    Our MDP is solved by the \emph{optimal} policy $\pi^\ast$ satisfying
    $$ \pi^\ast(s_i) = \argmax_\pi \E_{h_i|s_i,\pi}\left[ U(h_i) \right] \hspace{0.25in} \forall s_i \in S.$$
    We let $U^\ast(s_i){=} \E_{h_i|s_i,\pi^\ast}\left[ U(h_i) \right]$ denote the corresponding maximal value at each state $s_i{\in}S$.
\end{itemize}

Casting our problem into the MDP framework helps us design algorithmic solutions.
Our particular MDP lends itself to a straightforward dynamic programming approach, since there are no cycles in its directed graph of possible transitions.

\subsection{Solution via dynamic programming}
\label{sec:dp}

The MDP described in Section \ref{sec:formulation} can be solved by a relatively simple dynamic programming algorithm.
This makes our method a close relative of past dynamic programming approaches for trial design \citep{woodroofe_sequential_1990,hardwick_exact_1995, hardwick_induction_1999, hardwick_optimal_2002}.
However, our method differs from them in an important respect: 
we seek to maximize an objective that is a \emph{function of the trial history}, and not just a function of the final state.
Concretely, our objective function (Equation \ref{eq:utility}) includes $V(h)$ and $K(h)$, which are functions of block-wise attributes.
Formulating the problem as an MDP gives us the flexibility to consider such an objective.

\paragraph{Recurrence relations.} 
Like any dynamic programming algorithm, ours divides the problem at hand into subproblems and solves them in an order that efficiently reuses computation.
This dependence between subproblems is defined by a set of recurrence relations.
In our case we have a single recurrence based on the Bellman equation \citep{lattimore_bandit_2020}:
\begin{equation}
    U^\ast(s) = \begin{cases} 
                          ~\max_a \left\{ \E_{s^\prime \sim t(s, a)} \left[ R(s, a, s^\prime) + U^\ast(s^\prime) \right] \right\} & s \notin S_N \\[2em]
                          ~0 & s \in S_N
                      \end{cases}
                      \label{eq:recurrence}
\end{equation}
The algorithm computes this recurrence at every state in $S$, iterating through the state space in order of \emph{decreasing} $\underline{T}$.
In other words the algorithm evaluates the recurrence at each state in $S_{N}$, $S_{N-1}$, and so on, until it finally computes $U^\ast(s_0)$ for the the empty table $s_0 \in S_0$ and terminates. 
At each state $s \notin S_N$ the algorithm also tabulates the maximizing action $a^\ast$. 
This table of optimal actions is the algorithm's most important output, as it constitutes $\pi^\ast$, the optimized trial design.
Figure \ref{fig:pseudocode} illustrates the algorithm in detail with pseudocode.

The trial design (i.e., policy) yielded by this recurrence is guaranteed to maximize the expected utility (subject to the MDP formulation described in Section \ref{sec:formulation}), since our optimization problem has the \emph{optimal substructure property}.
See Appendix \ref{sec:optimal} of the Supplementary Materials for more discussion and a proof of optimal substructure.

\newcommand{\vnew}{$v_{\text{new}}$}
\begin{figure}
    \centering

    \begingroup
    \begin{minipage}{0.5\linewidth}
    \begin{algorithm}[H]
    \caption{\textsc{TrialMDP}}
    \begin{algorithmic}[1]
    \setlength{\baselineskip}{12pt}
    \Procedure{MainLoop}{$N, \lambda_F, \lambda_K$}
      \State initialize tables U, A
      \For{ $s \in S_{N}$ }
        \State U[s] = 0
      \EndFor
      \For{ $s \notin S_{N}$ }
        \State U[$s$] $ = -\infty$
        \For{ $a \in \mathcal{A}_s$ }
          \State $\text{u} = 0$
          \For{$(p, s^\prime) \in t(s, a)$}
            \State $\text{u} {\pluseq} p {\cdot} \big\{ R(a, s^\prime, N, \lambda_F, \lambda_K)$ 
            \State \hfill ${+} \text{U}[s^\prime] \big\}$
          \EndFor
          \If{u $>$ U[$s$]}
            \State U[s] = u; A[s] = $a$
          \EndIf
        \EndFor
      \EndFor
      \State \Return U, A
    \EndProcedure
    \end{algorithmic}
    \end{algorithm}
    \end{minipage}
    \begin{minipage}{0.45\linewidth}
    \begin{algorithmic}
    \setlength{\baselineskip}{12pt}
    \Function{$R$}{$a,s^\prime, N, \lambda_F, \lambda_K$}
        \State $r = 0$ 
        \State $w = a.N_A \cdot a.N_B / (a.N_A + a.N_B) $
        \State $\hat{p}_A = (s^\prime.n_A {+} 1)/ (s^\prime.N_A {+} 2)$
        \State $\hat{p}_B = (s^\prime.n_B {+} 1)/ (s^\prime.N_B {+} 2)$
        \State $\hat{q}_A = 1 - \hat{p}_A$
        \State $\hat{q}_B = 1 - \hat{p}_B$
        \State $r \pluseq 4 \cdot w / (N {\cdot} (\hat{p}_A {+} \hat{p}_B) (\hat{q}_A {+} \hat{q}_B))$
        \If{ $s^\prime \in S_N$ }
          \State $r \minuseq \lambda_F {\cdot} (s^\prime.N_A - s^\prime.N_B)$ 
          \State \hspace{0.5in} $\cdot(\hat{p}_B - \hat{p}_A)/N$
        \EndIf
        \State $r \minuseq \lambda_K$
      \State \Return $r$
     \EndFunction
     \end{algorithmic}
    \end{minipage}
    \endgroup
    
    \caption{\Ourmethod{} algorithm pseudocode. The algorithm populates tables U and A with optimal utilities and actions, respectively. 
    Tables U and A are indexed by states; i.e., U[$s$] yields the utility for state $s$. 
    The for-loop on line 5 iterates through states in order of decreasing $\underline{T}$.
    The for-loop on line 7 iterates through all possible actions for the current state;
    and the loop on line 9 computes the expectation of $U$ for the current state and action.
    Function $R$ evaluates the reward function given by Equation \ref{eq:reward}.
    We use ``dot notation'' to access the attributes of states and actions; 
    e.g., $s^\prime.N_A$ yields $N_A$ for state $s^\prime$.
    }
    \label{fig:pseudocode}
\end{figure}

\paragraph{Computational expense.} 
At a high level \Ourmethod{} is a nested loop over every possible state, action, and transition.
For each state the algorithm stores a set of values, along with the optimal action.
Hence the algorithm uses $O(|S|) {=} O(N^4)$ space.
The number of possible actions and transitions varies between states; summing across all states yields total time cost $O(|\Phi| N^7)$, where $\Phi$ is the set of allocation fractions mentioned in Section \ref{sec:formulation}.

These complexities apply if we allow the algorithm to consider every possible state and action.
However, there are practical ways to \emph{prune away} states and actions, attaining much lower computational cost without sacrificing much utility.
Introducing a \emph{minimum block size} parameter $T_{\text{min}}$ eliminates all of the states in $S_1, \ldots, S_{T_\text{min}-1}$ and $S_{N-T_{\text{min}}+1}, \ldots S_{N-1}$; and reduces the number of possible actions at each remaining state.
An additional \emph{block increment} parameter $\kappa$ further constrains the algorithm to states where $T$ is an integer multiple of $\kappa$, resulting in a ``coarsened'' state space.
These parameters reduce the algorithm's space and time cost to $O((N-T_{\text{min}})^4/ \kappa)$ and $O((N-T_{\text{min}})^7/\kappa^2)$, respectively.
See Appendix \ref{sec:complexity} of the Supplementary Materials for derivations.
We typically set $T_{\text{min}}{=}N/8$ and $\kappa{=}2$.
Unless specified otherwise, we use $\Phi = \{0.2, 0.3, \ldots, 0.8\}$.
These settings yielded trials with competitive characteristics, without incurring undue computational expense during the evaluations of Section \ref{sec:results}.

Empirically, we observe a time cost of 5; $2{,}300$; and $23{,}000$ seconds for trials with 40, 100, and 140 patients respectively.
These measurements used a single-threaded implementation of \Ourmethod{}, on a laptop with Intel 1.1GHz CPUs.

\section{Evaluation}
\label{sec:results}


\subsection{Simulation study}
\label{sec:sim_study}

We performed a simulation study to compare \Ourmethod{} against established trial designs.
At each point in a grid of values for $\lambda_F, \lambda_K, p_A,$ and $p_B$, we ran 10,000 simulated trials using \Ourmethod{} and a suite of baseline designs.
The baselines included \primo{} a 1:1, fixed randomization design;
\secundo{} a traditional Response-Adaptive Randomized (RAR) design; and
\tertio{} a blocked RAR design.

For null scenarios with $p_A {=} p_B$, we chose an arbitrary sample size of $N{=}100$.
For alternative scenarios with $p_A {>}p_B$, we chose $N$ large enough for a 1:1 design to attain a power of 0.8.
See Tables \ref{tab:alt_table} and \ref{tab:null_table} for the exact values of $N$, $p_A$, and $p_B$ used in our simulated scenarios.

The traditional RAR baseline used a 1:1 randomization ratio for the first $25\%$ of patients,
and adaptive randomization thereafter according to the procedure used by \cite{rosenberger_optimal_2001}.
That is, the $k^{\text{th}}$ patient was assigned to treatment $A$ with probability
\begin{equation}
\xi_k = \frac{\sqrt{\hat{\underline{p}}_{A,k{-}1}}}{\sqrt{\hat{\underline{p}}_{A,k{-}1}} + \sqrt{\hat{\underline{p}}_{B,k{-}1}}}.
\label{eq:rar_rule}
\end{equation}

The blocked RAR baseline used two blocks of equal size.
The first block used a 1:1 randomization ratio; the second block used the same randomization given by Equation \ref{eq:rar_rule}.
This agrees with the blocked RAR procedure described by \cite{chandereng_robust_2019}.

We used \Ourmethod{} to generate trial designs over a grid of parameter settings: $(\lambda_F, \lambda_K) \in \{2, 3, 4, 5\} {\times} \{0.01, 0.025, 0.05, 0.1\}$.
Each parameter setting implies a different balance between statistical power, patient outcomes, and the number of blocks.

\begin{figure}
    \centering
    \begin{minipage}{0.35\linewidth}
    \centering
    \includegraphics[width=\textwidth]{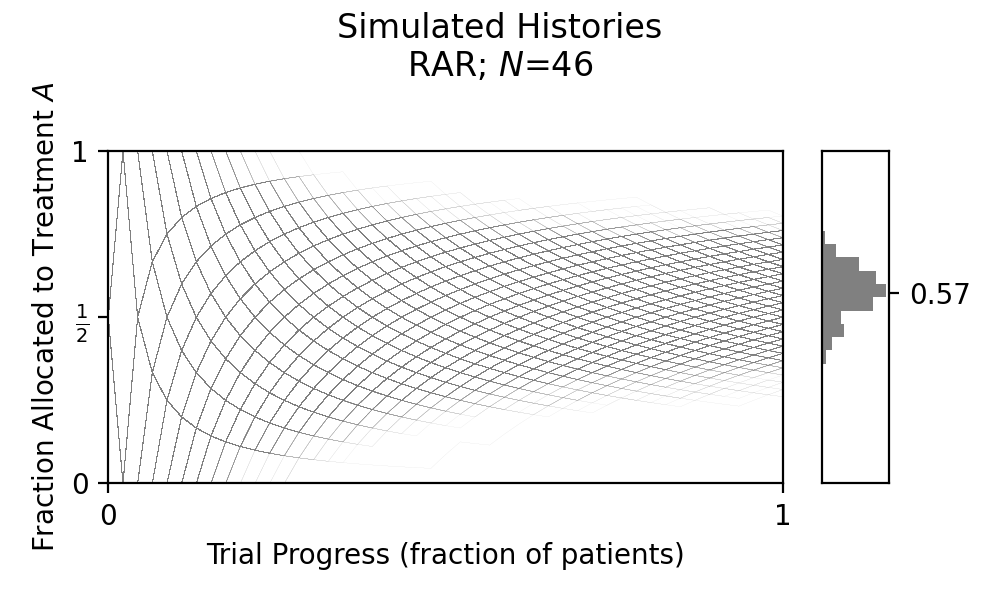}\\
    \includegraphics[width=\textwidth]{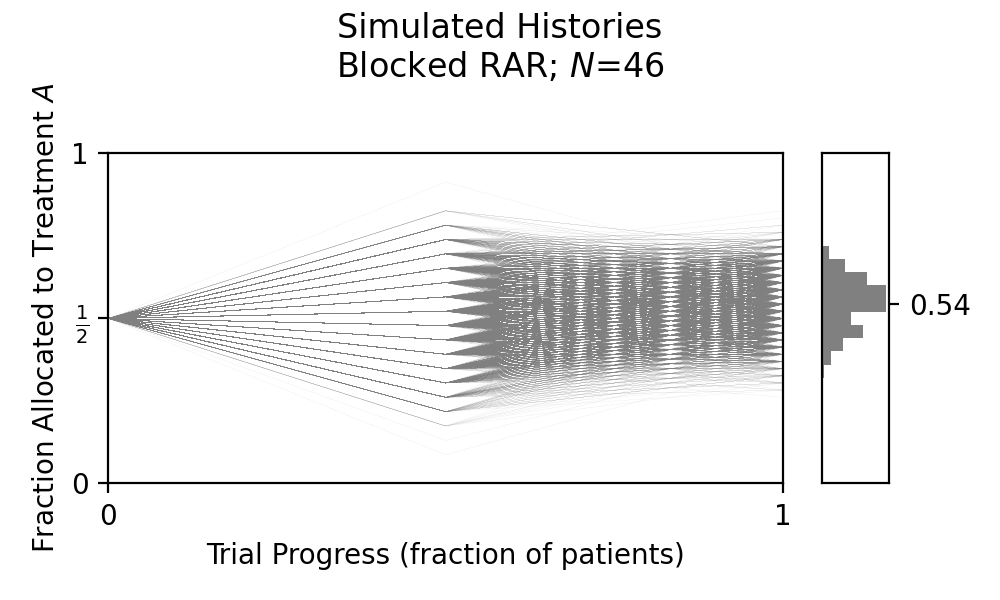}
    \end{minipage}%
    \begin{minipage}{0.64\linewidth}
    \centering
    \includegraphics[width=\textwidth]{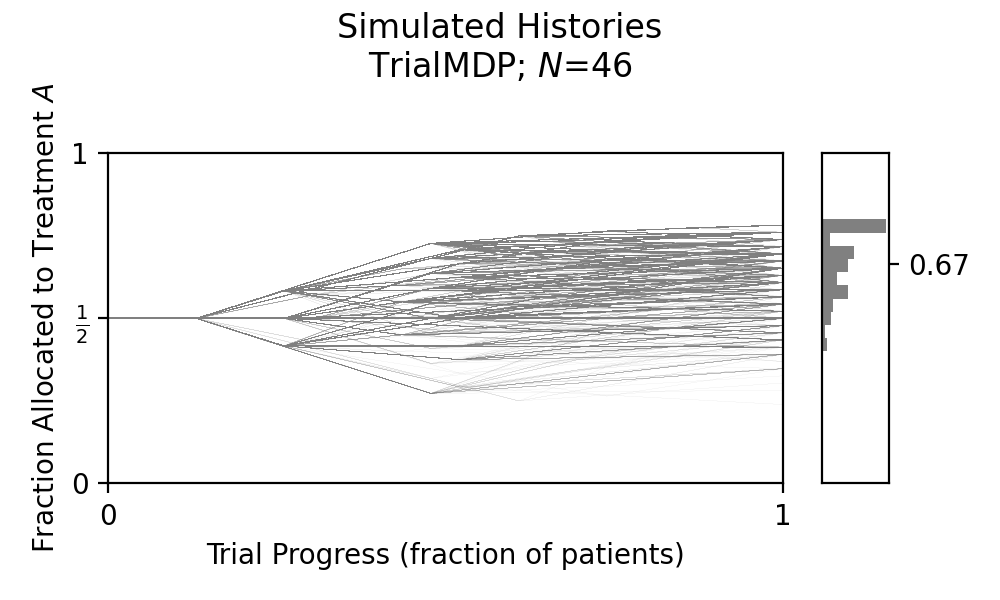}
    \end{minipage}\\
    \hspace{1.0in} (A) \hfill (B) \hspace{2.0in}
    \caption{Simulated trial histories. 
             Each plot traces the treatment allocation of 10,000 simulated trials.
             Histograms on the right give distributions of final allocations and report the mean.
             For each of these plots, $(p_A,p_B){=}(0.4, 0.1)$.
             (A) Histories for RAR and blocked RAR trial designs.
             (B) Histories for a \Ourmethod{} design, with parameter settings $\lambda_F{=}4.0$ and $\lambda_K{=}0.01$.
             Under these specific settings \Ourmethod{} allocates many more patients on average to the superior treatment.
             }
    \label{fig:histories}
\end{figure}

We simulated 10,000 trials for every scenario $p_A, p_B$, for each baseline design, and for each \Ourmethod{} parameter setting.
As an initial sanity check we visualized some trial histories to see whether the designs behaved as expected.
Figure \ref{fig:histories} shows some examples.
\Ourmethod{} always chose 1:1 allocation for the first block, increasing the allocation to $A$ in subsequent blocks when $p_A {>} p_B$. 
As $\lambda_F$ increased, the designs reliably increased allocation to $A$,
in agreement with our expectations.
The baseline designs also yielded trial histories that agreed with our expectations.

Recall that \Ourmethod{} is supposed to optimize the utility function (Equation \ref{eq:utility}) in expectation.
If this were true, we would expect our designs to attain higher utility than the others, averaged over the simulated histories.
To verify this we computed the utility for every simulated history and for every design, and tabulated the resulting averages.

Table \ref{tab:alt_table} shows some representative results from the alternative scenarios.
These results employed \Ourmethod{} with $\lambda_F{=}0.4$ and $\lambda_K{=}0.01$.
Under these particular parameter settings \Ourmethod{} attained slightly lower power than the other designs, but its superior patient outcomes gave it the greatest utility across all scenarios.
Indeed, we found that our algorithm does typically achieve higher average utility than the baseline designs \primo{} under the alternative hypothesis and \secundo{} \emph{as long as} $\lambda_F$ \emph{is sufficiently large}.
When $\lambda_F$ is not large enough, our designs have highest utility \emph{among the adaptive designs}, but the 1:1 design is mathematically guaranteed to attain highest utility.
We show this in Appendix \ref{sec:onestage} of the Supplementary Materials.

We highlight the fact that \Ourmethod{} assigned many more patients to the superior treatment on average, in all the scenarios of Table \ref{tab:alt_table}.
Furthermore, it did so reliably. 
The 5\%-ile for $N_A{-}N_B$ is higher for \Ourmethod{} than for any other adaptive design, in every alternative scenario.

It is also important to note that \Ourmethod{}'s design yielded slightly biased estimates of the effect size in the alternative scenarios.
We hypothesize that this bias---on the order of 0.01---stems from the rapidly changing randomization ratio prescribed by \Ourmethod{}.
The user ought to weigh this against other matters, such as vastly improved patient outcomes, when considering \Ourmethod{}.

Table \ref{tab:null_table} shows the corresponding results for null scenarios.
Notice that in some cases \Ourmethod{}'s designs showed somewhat inflated type-I error.
The percentiles of $N_A{-}N_B$ show that \Ourmethod{} is more prone to creating an imbalanced allocation under the null hypothesis.
Another salient observation is the relative decrease in utility for \emph{all} of the adaptive designs. 
This has a simple explanation. Under the null hypothesis, a 1:1 design always has optimal utility.
A 1:1 design attains maximal $V(h)$ and minimal $K(h)$; and under the null hypothesis, $F(h){=}0$ for any design.
Hence, every adaptive design will yield lower utility than the 1:1 design.

\begin{landscape}

\begin{table}[]
    \centering
    \resizebox{1.35\textwidth}{!}{
    \begin{tabular}{llr|rrrr|rrr|r@{\hskip 4pt}lr@{\hskip 4pt}lr@{\hskip 4pt}l|r|rrr}
\toprule
    &     &   & \multicolumn{4}{c}{Power (CMH test)} & \multicolumn{3}{c}{Effect Bias} & \multicolumn{6}{c}{$N_A {-} N_B$ (5\%, 95\%) } & \multicolumn{1}{c}{$K(h)$} & \multicolumn{3}{c}{Utility $Z$-score} \\
$p_A$ & $p_B$ & $N$ &           1:1 & RAR &  BRAR &   MDP &         RAR & BRAR & MDP &        \multicolumn{2}{c}{RAR} &  \multicolumn{2}{c}{BRAR} & \multicolumn{2}{c}{MDP} &                 MDP &            RAR & BRAR &  MDP  \\
\midrule
0.3 & 0.1 &   94 &           0.79 & 0.78 & 0.79 & 0.78 &            0.00 & 0.00 & 0.01 &      16.94 & (-4, 36)  & 11.22 & (-8, 30)  & 19.88 & (-2, 50) &              3.26 &            -60.25 & 6.23 &  8.40 \\
0.4 & 0.1 &   46 &           0.78 & 0.75 & 0.77 & 0.74 &            0.00 & 0.00 & 0.01 &      6.80  & (-6, 18)  &  3.98 & (-8, 16)  & 15.26 & (0, 26) &              3.87 &             -9.18 & 2.65 &  8.17 \\
    & 0.2 &  124 &           0.80 & 0.78 & 0.79 & 0.77 &            0.00 & 0.00 & 0.01 &      17.03 & (-6, 42)  & 11.18 & (-10, 32) & 31.77 & (-4, 66) &              3.50 &           -104.45 & 6.10 & 11.64 \\
0.5 & 0.3 &  144 &           0.80 & 0.80 & 0.78 & 0.76 &            0.00 & 0.00 & 0.01 &      14.20 & (-10, 38)  &  9.54 & (-12, 32) & 42.23 & (-6, 76) &              3.41 &           -157.52 & 5.11 & 14.80 \\
0.7 & 0.4 &   62 &           0.79 & 0.77 & 0.77 & 0.73 &            0.00 & 0.00 & 0.01 &      6.96  & (-8, 22)  &  4.60 & (-10, 18) & 23.03 & (0, 34) &              3.85 &            -17.25 & 2.89 & 11.22 \\
    & 0.5 &  144 &           0.81 & 0.79 & 0.79 & 0.77 &            0.00 & 0.00 & 0.01 &      9.22  & (-12, 30) &  6.13 & (-14, 28) & 42.33 & (-6, 76) &              3.34 &           -174.98 & 3.16 & 14.78 \\
0.9 & 0.6 &   46 &           0.79 & 0.78 & 0.78 & 0.76 &            0.00 & 0.00 & 0.01 &      3.72  & (-8, 16)  &  2.55 & (-8, 14)  & 13.43 & (0, 26) &              3.50 &             -9.56 & 1.62 &  6.89 \\
    & 0.7 &   94 &           0.80 & 0.80 & 0.80 & 0.79 &            0.00 & 0.00 & 0.00 &      4.53  & (-12, 20) &  3.08 & (-14, 20) & 19.17 & (-2, 50) &              3.22 &            -73.27 & 1.32 &  7.80 \\
\bottomrule
\end{tabular}
   
    }
    \caption{Simulation study alternative scenarios.
             Labels RAR, BRAR, and MDP refer to adaptive trials designed by traditional RAR, blocked RAR, and \Ourmethod{}, respectively.
             The label 1:1 refers to a fixed randomization trial with one-to-one allocation.
             The ``Effect Bias'' multicolumn reports the average difference between estimated effect size and true effect size.
             The ``$N_A {-} N_B$ (5\%, 95\%)'' multicolumn shows the difference in patient allocation between treatments; 
             it reports the mean, with the 5\%-ile and 95\%-ile in parentheses.
             $K(h)$ shows the average number of blocks. It only varies for \Ourmethod{}; $K(h){=}N$ for RAR and $K(h){=}2$ for BRAR in all scenarios. 
             The ``Utility $Z$-score'' multicolumn reports gain in utility relative to the 1:1 trial design, computed as $Z{=}(\mu_1 - \mu_2)/\sqrt{(\sigma_1^2 + \sigma_2^2)/10{,}000}$.
             For these results, \Ourmethod{} used parameter settings $\lambda_F{=}4.0$ and $\lambda_K{=}0.01$.
              }
    \label{tab:alt_table}
\end{table}

\begin{table}[]
    \centering
    \resizebox{1.35\textwidth}{!}{
    \begin{tabular}{lr|rrrr|rrr|r@{\hskip 4pt}lr@{\hskip 4pt}lr@{\hskip 4pt}l|r|rrr}
\toprule
    &      & \multicolumn{4}{c}{Size (CMH test)} & \multicolumn{3}{c}{Effect Bias} & \multicolumn{6}{c}{$N_A{-}N_B$ (5\%, 95\%)} & \multicolumn{1}{c}{$K(h)$} & \multicolumn{3}{c}{Utility $Z$-score} \\
$p_A{=}p_B$ &  $N$ &    1:1 & RAR &  BRAR & MDP &        RAR & BRAR & MDP &           \multicolumn{2}{c}{RAR} & \multicolumn{2}{c}{BRAR} & \multicolumn{2}{c}{MDP} &            MDP &         RAR &  BRAR &    MDP \\
\midrule
0.1 &  100 &          0.05 & 0.05 & 0.05 & 0.05 &       0.00 & 0.00 & 0.00 &   0.25 & (-22, 24) &  0.14 & (-20, 22) &  0.13 & (-18, 18) &       2.77 &    -542.40 & -5.26 &  -9.15 \\
0.3 &  100 &          0.05 & 0.05 & 0.05 & 0.05 &       0.00 & 0.00 & 0.00 &  -0.18 & (-22, 20) & -0.07 & (-20, 20) & -0.07 & (-32, 32) &      3.77 &     -505.00 & -5.14 & -13.45 \\
0.5 &  100 &          0.05 & 0.05 & 0.05 & 0.06 &       0.00 & 0.00 & 0.00 &   0.05 & (-18, 18) &  0.12 & (-18, 18) &  0.06 & (-46, 46) &      3.88 &     -490.50 & -4.96 & -13.36 \\
0.6 &  100 &          0.05 & 0.05 & 0.04 & 0.05 &       0.00 & 0.00 & 0.00 &   0.03 & (-18, 18) &  0.07 & (-18, 18) & -0.04 & (-42, 42) &       3.85 &    -491.29 & -4.98 & -13.31 \\
0.7 &  100 &          0.05 & 0.05 & 0.05 & 0.06 &       0.00 & 0.00 & 0.00 &   0.11 & (-18, 18) & -0.01 & (-16, 16) &  0.32 & (-36, 36) &       3.68 &    -499.43 & -5.12 & -12.64 \\
0.9 &  100 &          0.05 & 0.05 & 0.05 & 0.05 &       0.00 & 0.00 & 0.00 &  -0.10 & (-16, 16) & -0.01 & (-16, 16) & -0.02 & (-18, 18) &       2.77 &    -511.70 & -5.05 &  -9.03 \\
\bottomrule
\end{tabular}

    }
    \caption{Simulation study null scenarios.
             We report trial size rather than power;
             all other columns have the same meaning as in Table \ref{tab:alt_table}.
             For these results, \Ourmethod{} used $\lambda_F{=}4.0$ and $\lambda_K{=}0.01$.}
    \label{tab:null_table}
\end{table}

\begin{table}[]
    \centering
    \resizebox{1.35\textwidth}{!}{
    \begin{tabular}{llr|rrrr|rrr|r@{\hskip 4pt}lr@{\hskip 4pt}lr@{\hskip 4pt}l|r|rrr}
\toprule
    &     &   & \multicolumn{4}{c}{Power/Size (CMH test)} & \multicolumn{3}{c}{Effect Bias} & \multicolumn{6}{c}{$N_A {-} N_B$ (5\%, 95\%) } & \multicolumn{1}{c}{$K(h)$} & \multicolumn{3}{c}{Utility $Z$-score} \\
$p_A$ & $p_B$ & $N$ &           1:1 & RAR &  BRAR &   MDP &         RAR & BRAR & MDP &        \multicolumn{2}{c}{RAR} &  \multicolumn{2}{c}{BRAR} & \multicolumn{2}{c}{MDP} &                 MDP &            RAR & BRAR &  MDP  \\
\midrule
0.4 & 0.4 &   20 &           0.06 & 0.05 & 0.05 & 0.05 &            0.00 & 0.00 & 0.00 &      -0.04 & (-8, 8)   & -0.03 & (-8, 8)   & 0.01  & (-10, 10)&              2.57 &            -45.59 & -2.48 & -3.57 \\ 
0.8 & 0.4 &   20 &           0.61 & 0.60 & 0.54 & 0.56 &            0.00 & 0.00 & 0.02 &       2.13 & (-6, 10) &  1.47  & (-6, 8)   & 5.09  & (-6, 10) &              2.28 &             -9.94 &  0.24 &  2.50 \\
\bottomrule
\end{tabular}


    }
    \caption{Results of trial redesign. 
             We report the same quantities as in the simulation study.
             For these results, \Ourmethod{} used $\lambda_F{=}3.0$ and $\lambda_K{=}0.05$.}
    \label{tab:redesign}
\end{table}

\end{landscape}

\begin{figure}
    \centering
    \includegraphics[scale=0.45]{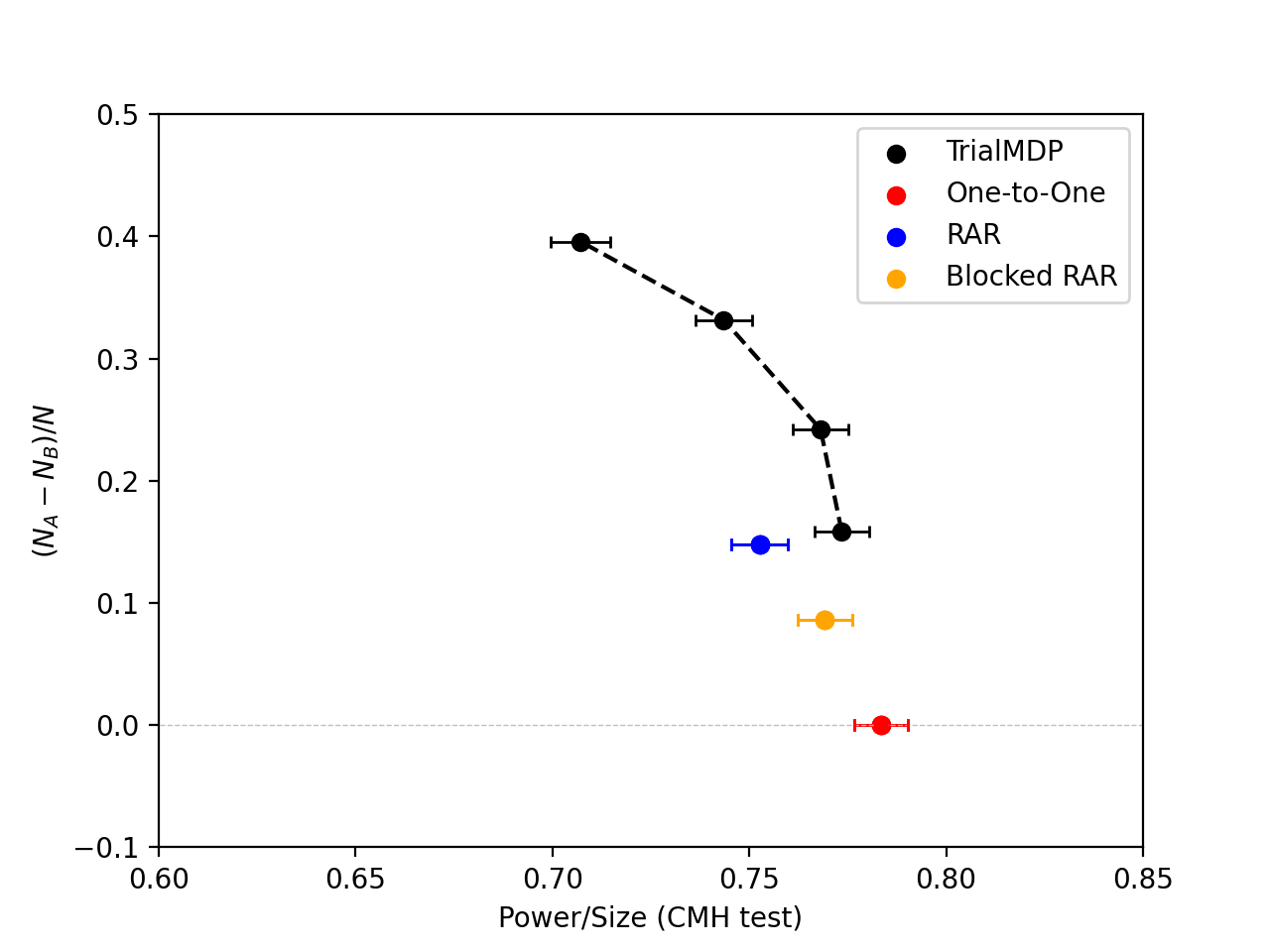}
    \includegraphics[scale=0.45]{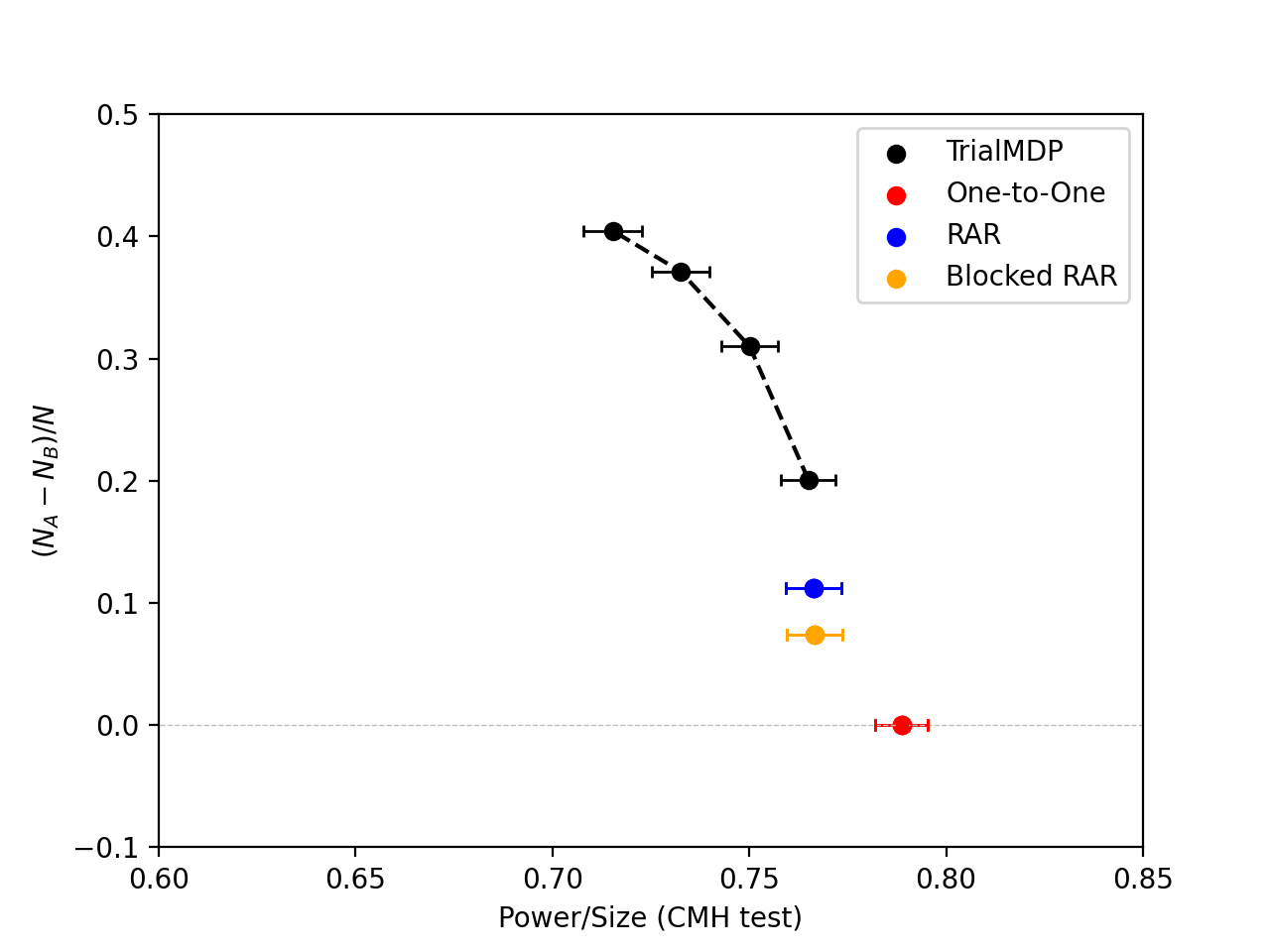}\\[-1em]
    (A) \hspace{2.5in} (B)
    \caption{Frontier plots. 
    (A) Trial design characteristics for scenario $p_A{=}0.4, p_B{=}0.1$.
    (B) Trial design characteristics for scenario $p_A{=}0.7, p_B{=}0.4$
    In each plot the curve traced by \Ourmethod{} corresponds to $\lambda_F{\in}\{2,3,4,5\}$.
    \Ourmethod{} used $\lambda_K{=}0.01$ in both plots.
    The points represent \emph{mean} power and patient allocations;
    the error bars show symmetric 90\% confidence intervals for the means.
    Note that there are error bars for the vertical direction, but they are too compact to be seen.
    }
    \label{fig:frontiers}
\end{figure}

Beyond a one-dimensional comparison of utility, it is useful to compare the designs in two dimensions: statistical power and patient outcomes.
As we vary the parameter $\lambda_F$, \Ourmethod{} designs trials that balance these quantities differently.
We visualize this with \emph{frontier plots}; 
trial designs are shown as points in two dimensions, with statistical power on the horizontal axis and allocation to $A$ on the vertical axis.
Higher and to the right is better.
Figure \ref{fig:frontiers} gives examples.
In some scenarios, \Ourmethod{}'s designs \emph{dominate} the other adaptive designs, attaining higher power \emph{and} assigning more patients to treatment $A$.
Figure \ref{fig:frontiers}(A) is one such case.
In other scenarios, \Ourmethod{}'s designs do not dominate the others. 
However, Figure \ref{fig:frontiers}(B) shows that even in those cases, \Ourmethod{} still provides a useful way to control the balance between power and patient outcomes.
For example, the user is free to choose a trial design with much better patient outcomes at the cost of slightly lower power, by selecting larger values of $\lambda_F$.

\subsection{Trial redesign}
\label{sec:redesign}

We demonstrate \Ourmethod{}'s practical usage by applying it to a historical trial.
We chose the phase-II thymoglobulin trial described by \cite{bashir_randomized_2012} because it
\primo{} had two arms,
\secundo{} had a small sample size ($N{=}20$), and
\tertio{} the trial designers saw fit to use an adaptive design, for ethical reasons.
This combination made the trial well-suited for testing our algorithm.

We redesigned the trial in two phases: ``parameter tuning'' and ``testing.''
In the parameter tuning phase we swept through the same grid of $\lambda_F, \lambda_K, p_A, p_B$ values used in our simulation study, but with the sample size fixed at $N{=}20$.
We ran our algorithm and simulated 10,000 trials at each grid point, and generated frontier plots similar to those in Figure \ref{fig:frontiers}.
Visual inspection suggested that \Ourmethod{} with $\lambda_F{=}3.0$ and $\lambda_K{=}0.05$ would yield reasonable power and patient outcomes for a variety of $p_A, p_B$ scenarios.

In the testing phase we simulated the thymoglobulin trial by computing point estimates of $p_A{=}0.8$ and $p_B{=}0.4$ from the original trial's results. 
We simulated two scenarios: a null scenario where $p_A{=}p_B{=}0.4$, and an alternative where $p_A{=}0.8$ and $p_B{=}0.4$. 
Using the design from \Ourmethod{} with ``tuned'' parameter values $\lambda_F{=}3.0$ and $\lambda_K{=}0.05$, we simulated 10,000 trials for each scenario. 
The results are aggregated in Table \ref{tab:redesign}.
Under the alternative scenario we found that \Ourmethod{}'s design, on average, assigned significantly more patients to treatment $A$ with a slightly decreased power of 0.557.
Note also that in the null scenario, \Ourmethod{}'s design had a somewhat inflated type-I error of 0.055.

\section{Discussion}
\label{sec:discussion}

\paragraph{Key takeaways.}
We presented \Ourmethod{}, an algorithm for designing blocked RAR trials.
\Ourmethod{} represents a blocked RAR trial as a Markov Decision Process, and solves for the optimal design via dynamic programming.
The resulting design dictates the size and treatment allocation of the next block, given the results observed thus far.

Our algorithm allows users to choose the relative importance of \primo{} statistical power and \secundo{} patient outcomes.
The trial designs generated by \Ourmethod{} consistently attain superior utility against a suite of baselines when \primo{} the effect size is large and \secundo{} patient outcomes are given sufficient importance.
The simulation study in Section \ref{sec:sim_study} demonstrates this.

\Ourmethod{} has some shortcomings worth keeping in mind.
It is currently restricted to a narrow class of trials: two-armed trials with binary outcomes.
All outcomes for past blocks must be observed before the next block can begin.
The MDP formulation assumes a single statistical test (one-sided CMH) is performed at the end of the trial.
While interim analyses may be used in trials governed by the current version of \Ourmethod{}, we provide no guarantees of optimality in that case.
\Ourmethod{}'s computational cost grows quickly with the number of patients, and becomes impractical for $N{>}200$.
Setting large values for the \emph{minimum block size} and \emph{block increment} parameters ($T_{\text{min}}$ and $\kappa$) can ameliorate some of this expense. 
Simulations showed that in some scenarios, \Ourmethod{}'s designs have modestly inflated type-I error, and may yield a slightly biased estimate of effect size.
These weaknesses should be weighed against the vastly superior patient outcomes \Ourmethod{} can deliver.

\paragraph{Practical recommendations.}
The user of \Ourmethod{} immediately faces a question: \emph{what values of $\lambda_F$ and $\lambda_K$ should be used?}
Consider the terms of Equation \ref{eq:utility}.
Since $V(h)$ is only a proxy for the statistical power, there isn't a clear way to assign practical meaning to $\lambda_F, \lambda_K$.
For example, we cannot interpret $\lambda_F$ as a literal ``conversion rate'' between units of failure and units of statistical power.
This makes it difficult to set $\lambda_F, \lambda_K$ in a principled way.
Instead we recommend \emph{tuning} $\lambda_F$ and $\lambda_K$ through a process like the one demonstrated in Section \ref{sec:redesign}:
\primo{} use the algorithm to design trials for a grid of $\lambda_F, \lambda_K$ values;
\secundo{} simulate trials for each design, for a set of scenarios $p_A, p_B$; 
\tertio{} examine the simulation results and choose $\lambda_F, \lambda_K$ that yield acceptable power and patient outcomes across scenarios.
As a starting point, $\lambda_F{=}3.0$, $\lambda_K{=}0.01$ yielded reasonable characteristics across all the scenarios in this paper.


\paragraph{Future improvements.}
Although \Ourmethod{}'s current implementation is single-threaded, it is highly parallelizable and would have a speedup roughly linear in the number of threads.
A multi-threaded parallel implementation is a natural next step.

There are multiple ways that \Ourmethod{} could be extended to a broader class of trials.
For instance, it could permit more than two arms and more than two outcomes. 
This would incur exponentially greater computational expense, but may be useful for some very small trials.

The current MDP formulation assumes that the trial terminates after all patients have been treated.
A more sophisticated MDP could incorporate interim analyses, accounting for the possibility of early termination for success or futility.

\section{Software}
\label{sec:software}

We implemented \Ourmethod{} in C++ and provide it as an R package on GitHub:\\ 
\Githubrepo{}.
We also provide the code for our Section \ref{sec:results} evaluations in another repository: \Analysisrepo{}. 
This includes a Snakemake workflow \citep{molder_snakemake_2021} that reproduces all results in this paper.

\section{Supplementary Material}
\label{sec6}
Our Supplementary Material contains four appendices. 
Appendix A gives our justification for using the function $V(h)$.
Appendix B shows that our optimization problem has the optimal substructure property (and hence \Ourmethod{} yields an optimal policy with respect to our MDP assumptions).
Appendix C derives the computational complexities given in Section \ref{sec:dp}.
Appendix D shows that $\lambda_F$ must be sufficiently large for an adaptive trial to attain higher utility than a single-block trial.

\section*{Acknowledgments}

We thank Zhu Xiaojin and Blake Mason for conversations about bandit algorithms.
DM was funded by the National Institutes of Health (award T32LM012413).

{\it Conflict of Interest}: None declared.

\bibliographystyle{biorefs}
\bibliography{refs}

\newpage

\clearpage

\appendix

\section{Derivation of $V$}
\label{sec:v_score}

Equation \ref{eq:utility} uses the following function, $V(h)$, as a proxy for a trial's statistical power:
\begin{equation*}
    V(h) = \frac{1}{N} \sum_i \frac{w_i}{\frac{1}{2}(\underline{p}_{A,i} + \underline{p}_{B,i})\frac{1}{2}(\underline{q}_{A,i} + \underline{q}_{B,i})}
\end{equation*}
where 
$$w_i = N_{A,i} {\cdot} N_{B,i} / (N_{A,i} {+} N_{B,i}).$$ 
This appendix provides some justification for $V(h)$.

We're interested in blocked RAR trials where the final analysis uses a Cochran-Mantel-Haenzsel (CMH) superiority test.
Recall that the CMH statistic takes this form:
$$ \CMH(h) =  \frac{\sum_i w_i d_i}{\sqrt{ \sum_i w_i \hat{p}_i \hat{q}_i} }$$
where 
$$
d_i = p_{A,i} - p_{B,i}
\hspace{0.5in}
\hat{p}_i = \frac{N_{A,i} {\cdot} p_{A,i} + N_{B,i} {\cdot} p_{B,i}}{T_i}
\hspace{0.5in} 
\hat{q}_i = 1 - \hat{p_i}
$$
Under the null hypothesis, $\CMH \sim \mathcal{N}(0,1)$ asymptotically.
Intuitively, we maximize the power of the test by choosing $N_{A,i}, N_{B,i}$ such that when $p_A {\neq} p_B$, the distribution of $\CMH$ has large mean without inflated variance.
Our goal is to find an objective function $V$ that, when maximized, yields trial designs with those characteristics.

As a first candidate we may try maximizing the expected value of of $\CMH$:
\begin{align*}
    \E_h \left[\CMH(h)\right] &\simeq \frac{\sum_i w_i (p_A - p_B)}{\sqrt{\sum_i w_i \left( \frac{N_{A,i} {\cdot} p_A + N_{B,i} {\cdot} p_B}{T_i} \right) \left( \frac{N_{A,i} {\cdot} q_A + N_{B,i} {\cdot} q_B}{T_i} \right) }} \\
    &= (p_A - p_B) \frac{\sum_i w_i}{\sqrt{\sum_i w_i \left(\phi p_A + (1-\phi) p_B \right) \left( \phi q_A + (1-\phi) q_B\right) } } 
\end{align*}
where $\phi_i = N_{A,i} / T_i$ is the \emph{fraction} of block $i$'s patients allocated to $A$.
The trial designer has no control over $p_A$ or $p_B$.
So if they wish to maximize this quantity then they may ignore the factor $(p_A - p_B)$, yielding
\begin{equation}
    \frac{\sum_i w_i}{\sqrt{\sum_i w_i \left(\phi p_A + (1-\phi) p_B \right) \left( \phi q_A + (1-\phi) q_B\right) } }
    \label{eq:expected-cmh}
\end{equation}
as a proxy objective for maximizing power.
It's important to note, however, a subtle property of Expression \ref{eq:expected-cmh}.
The denominator is minimized when more patients are allocated to the treatment with more \emph{extreme} success probability---i.e., success probability closer to 0 or 1.
As a result, the maximizer of Expression \ref{eq:expected-cmh} exhibits a preference toward that treatment.
This preference manifested itself in earlier versions of the algorithm, which would do well when $\frac{1}{2} < p_B < p_A$, but would do worse when $p_B < p_A < \frac{1}{2}$.

As a second candidate, we may try maximizing the the related quantity
\begin{align}
    \frac{\E\left[ \sum_i w_i d_i \right]}{\sqrt{\text{Var} \left[ \sum_i w_i d_i \right]}} &= (p_A - p_B) \frac{\sum_i w_i }{\sqrt{\sum_i w_i^2 \left( \frac{p_A q_A}{N_{A,i}} + \frac{p_B q_B}{N_{B,i}} \right)}} \label{eq:cmh_mean_div_std} \\[1em]
    &= (p_A - p_B) \frac{\sum_i w_i }{ \sqrt{ \sum_i w_i \left( (1{-}\phi_i)p_A q_A + \phi_i p_B q_B \right)} } \\[1em]
    &\propto \frac{\sum_i w_i }{ \sqrt{ \sum_i w_i \left( (1{-}\phi_i)p_A q_A + \phi_i p_B q_B \right)} }
    \label{eq:hm-v2}
\end{align}
\citeauthor{cochran_cmh_1954} uses Expression \ref{eq:cmh_mean_div_std} as a proxy for the power of a CMH test in his original justifications for the CMH statistic \citep{cochran_cmh_1954}.
Like Expression \ref{eq:expected-cmh}, the new Expression \ref{eq:hm-v2} also exhibits a preference based on \emph{extremality} of the success probabilities. 
However, it instead favors the treatment with \emph{less} extreme success probability, i.e., probability nearer $\frac{1}{2}$.
Versions of the algorithm based on Expression \ref{eq:hm-v2} would manifest this preference during simulations.
The algorithm would attain superior utility when $p_B < p_A < \frac{1}{2}$, but would do worse when $\frac{1}{2} < p_B < p_A$.

Note the similarity between Expression \ref{eq:hm-v2} and Expression \ref{eq:expected-cmh}.
They have identical numerators, and both denominators have the form $\sqrt{\sum_i w_i \tilde{pq}}$ where $\tilde{pq}$ is some ``combined variance'' computed from $p_A, p_B$.
They differ precisely in how they compute $\tilde{pq}$. 
This in turn produces their different preferences (toward the treatment with less-extreme and more-extreme success probability, respectively).
Neither of these preferences are favorable.
We would like a proxy for power that has simpler dependence on $p_A$ and $p_B$, which are unknown.
To that end we propose our final candidate:
\begin{align*}
    \frac{\sum_i w_i}{\sqrt{\sum_i w_i \frac{1}{2}(p_{A} + p_{B})\frac{1}{2}(q_{A} + q_{B})}} 
    &= \frac{\sum_i w_i}{\sqrt{\frac{1}{2}(p_{A} + p_{B})\frac{1}{2}(q_{A} + q_{B})} \sqrt{\sum_i w_i }} \\[1em]
    &= \sqrt{ \frac{\sum_i w_i}{\frac{1}{2}(p_{A} + p_{B})\frac{1}{2}(q_{A} + q_{B})}}
\end{align*}
or, after squaring,
\begin{equation*}
    \frac{\sum_i w_i}{\frac{1}{2}(p_{A} + p_{B})\frac{1}{2}(q_{A} + q_{B})}
    \label{eq:current_w}
\end{equation*}
This new quantity lets $\tilde{pq} = \frac{1}{2}(p_{A} + p_{B})\frac{1}{2}(q_{A} + q_{B})$, which has neither of the preferences exhibited by Expressions \ref{eq:expected-cmh} or \ref{eq:hm-v2}.
Of course in practice we don't know $p_A$ or $p_B$, so we substitute their MAP estimates at each block:
\begin{equation}
    V(h) = \sum_i \frac{w_i}{\frac{1}{2}(\underline{\hat{p}}_{A,i} + \underline{\hat{p}}_{B,i})\frac{1}{2}(\underline{\hat{q}}_{A,i} + \underline{\hat{q}}_{B,i})},
\end{equation}
which is the expression for $V$ used in Section \ref{sec:formulation} (up to a factor of $\frac{1}{N}$).

\section{Optimal Substructure}
\label{sec:optimal}
We show that our optimization problem---maximizing expected utility---possesses the \emph{optimal substructure} property.
In other words, we prove that the recurrence relation (Equation \ref{eq:recurrence}) correctly decomposes the problem into subproblems, and reuses their solutions to solve the original problem.

Suppose the algorithm is evaluating $U^\ast(s_i)$ for some state $s_i$, and that it's already evaluated $U^\ast(s_{i+1})$ for every possible successor state $s_{i+1}$ of $s_i$.
Let $\pi^\ast$ denote the optimal policy, i.e., the one yielding $U^\ast$.
Then optimal substructure follows from the linearity of our utility function.
Assuming $s_{i+1}$ is not terminal:
\begin{align*}
    U^\ast(s_i) &= \E_{h_{i+1}|\pi^\ast, s_i} \left[\frac{1}{N}V(h_{i+1}) - \lambda_F {\cdot} F(h_{i+1}) -\lambda_K {\cdot} K(h_{i+1}) \right] \\[1em]
                   &= \E_{(s_{i+1}, h_{i+2}) | \pi^\ast, s_i} \left[\frac{w(a^\ast)}{N \tilde{pq}(s_{i+1}) } + \frac{1}{N}V(h_{i+2}) - \lambda_F {\cdot} F(h_{i+2}) -\lambda_K {\cdot}( 1+  K(h_{i+2}) ) \right] \\[1em]
                   &= \E_{s_{i+1} | \pi^\ast, s_i} \left[ \E_{h_{i+2}|\pi^\ast, s_{i+1}} \left[ \frac{w(a^\ast)}{N \tilde{pq}(s_{i+1}) } + \frac{1}{N}V(h_{i+2}) - \lambda_F {\cdot} F(h_{i+2}) -\lambda_K {\cdot}( 1+  K(h_{i+2}) )\right] \right] \\[1em]
                   &= \E_{s_{i+1} | \pi^\ast, s_i} \left[  \frac{w(a^\ast)}{N \tilde{pq}(s_{i+1}) } - \lambda_K + \E_{h_{i+2}|\pi^\ast, s_{i+1}} \left[\frac{1}{N}V(h_{i+2}) - \lambda_F {\cdot} F(h_{i+2}) -\lambda_K {\cdot}K(h_{i+2} )\right] \right] \\[1em]
                   &= \E_{s_{i+1} | \pi^\ast, s_i} \left[  R(a^\ast, s_{i+1}) + U^\ast(s_{i+1}) \right] \\[1em]
                   &= \max_a \left\{ \E_{s_{i+1} |  s_i} \left[  R(a, s_{i+1}) + U^\ast(s_{i+1}) \right] \right\}
\end{align*}
which agrees exactly with the recurrence in Equation \ref{eq:recurrence}. 
A similar computation covers the case when $s_{i+1}$ is terminal.

Put another way, our dynamic program's recurrence relation computes $U^\ast(s)$ correctly at each state, and will yield the optimal policy $\pi^\ast$.

\section{Computational complexity}
\label{sec:complexity}

We derive the space and time complexities given in Section \ref{sec:dp}.

Let $S_i$ denote the set of all $2{\times}2$ contingency tables containing $i$ observations.
Define $[\tmin{}:\kappa:N-\tmin{}] = \{\tmin{}, \tmin+\kappa, \ldots, N-\tmin{} \}$, the set of integers ranging from $\tmin{}$ to $N-\tmin{}$ in increments of $\kappa$.
Then $|S_i| = O(i^3)$, and the size of the full state space is 
$$|S| ~=~ \sum_{i \in [\tmin{}:\kappa:N-\tmin{}]} |S_i| ~=~ \sum_{i=\tmin{}}^{N-\tmin{}} O\left(\frac{i^3}{\kappa} \right) ~=~ O\left(\frac{(N-\tmin{})^4}{\kappa} \right).$$ 
The algorithm stores data proportional to $|S|$, so this gives the space complexity.

The time complexity results from a nested sum over states, actions, and transitions:
\begin{align*}
    T(N, |\Phi|) &= \sum_{i \in [\tmin{}:\kappa:N{-}\tmin{}]} ~ \sum_{s \in S_i} ~ \sum_{j \in [\tmin{}:\kappa:N{-}\tmin{}-i]} ~ \sum_{\phi \in \Phi} \phi{\cdot} j (1 - \phi){\cdot} j \\[1em]
         &= \sum_{i \in [\tmin{}:\kappa:N{-}\tmin{}]} ~ \sum_{s \in S_i} ~ \sum_{j \in [\tmin{}:\kappa:N{-}\tmin{}-i]} ~ j^2 \sum_{\phi \in \Phi} \phi (1 - \phi) \\[1em] 
         &= \sum_{i \in [\tmin{}:\kappa:N{-}\tmin{}]} ~ \sum_{s \in S_i} ~ \sum_{j \in [\tmin{}:\kappa:N{-}\tmin{}-i]} j^2 {\cdot} O(|\Phi|) \\[1em]
         &= O(|\Phi|) \cdot \sum_{i \in [\tmin{}:\kappa:N{-}\tmin{}]} ~ \sum_{s \in S_i} ~ \sum_{j \in [\tmin{}:\kappa:N{-}\tmin{}-i]} j^2 \\[1em]
         &= O(|\Phi|) \cdot \sum_{i \in [\tmin{}:\kappa:N{-}\tmin{}]} O(i^3)\sum_{j \in [\tmin{}:\kappa:N{-}\tmin{}-i]} j^2 \\[1em]
         &= O(|\Phi|) \cdot \sum_{i \in [\tmin{}:\kappa:N{-}\tmin{}]} O(i^3) O\left(\frac{(N-\tmin{}-i)^3}{\kappa}\right) \\[1em]
         &= O(|\Phi|) \cdot O\left(\frac{(N-\tmin{})^3}{\kappa}\right) \cdot \sum_{i \in [\tmin{}:\kappa:N{-}\tmin{}]} O(i^3) \\
         &= O\left(|\Phi| \cdot \frac{(N - \tmin{})^7}{\kappa^2}\right).
\end{align*}

\section{The utility of single-block vs. multi-block trials}
\label{sec:onestage}
It is not always possible for a trial with multiple blocks to attain higher utility (Equation \ref{eq:utility}) than a trial with one block.
The cost $\lambda_K$ of an additional block outweighs any improvements in patient outcomes, unless $\lambda_F$ is large enough.
In this appendix we find conditions on $\lambda_F, \lambda_K$ that determine when a two-block adaptive trial can attain higher utility than a single-block (i.e., fixed randomization) trial.

Given the true values of $p_A$ and $p_B$, we can compute the utility of a single-block trial in closed form:
$$ U_{\text{single}} = \frac{1}{(p_A + p_B)(q_A + q_B)} - \lambda_K$$
We can likewise compute the utility of a two-block trial in closed form.
Assume the first block of the trial treats $T$ patients, assigning half to each treatment.
In the second block, assume $\phi {\cdot} (N-T)$ patients are assigned to treatment $A$ and $(1-\phi) {\cdot} (N-T)$ are assigned to treatment $B$.
Then the two-block trial has this utility:
$$ U_{\text{two-block}} = \frac{T + (N-T)\phi(1-\phi)}{N(p_A + p_B)(q_a + q_B)} + \frac{\lambda_F}{N}(2 \phi - 1)(N-T)(p_A-p_B) - 2\lambda_K $$

We want to find conditions where $U_{\text{two-block}} - U_{\text{single}} \geq 0$.
Some algebra yields this condition:
$$-\frac{1}{(p_A + p_B)(q_A + q_B)} \hat{\phi}^2  + \lambda_F (p_A - p_B) \hat{\phi} - \lambda_B \frac{N}{N-T} \geq 0$$
where $\hat{\phi} = (2\phi - 1)$ is a convenient shorthand.
The LHS of this inequality is a concave quadratic in $\hat{\phi}$.
It has real roots (and hence, a feasible region) iff
$$ \lambda_F^2 (p_A - p_B)^2 (p_A + p_B)^2 (q_A + q_B)^2 - 4 \frac{\lambda_B (p_A + p_B) (q_A + q_B) N}{N-T} \geq 0.$$
Rearranging gives the following condition on $\lambda_F$
$$ \lambda_F \geq \frac{2}{(p_A - p_B)} \sqrt{ \frac{N \lambda_K}{(N-T) (p_A + p_B) (q_A + q_B)} }.$$

The key takeaway is that $\lambda_F$ must be sufficiently large before \emph{any} RAR design can possibly attain higher utility than a single-block, fixed-randomization trial.

This analysis does not account for the uncertainty in $p_A$ and $p_B$. 
\Ourmethod{} operates under this uncertainty, and will therefore not generally choose a single-block design when it is truly optimal.
In practice, the algorithm only chooses a single-block design when $\lambda_F$ is small relative to $\lambda_K$.


\end{document}